\documentclass[letterpaper]{article}

\usepackage{float}
\usepackage{latexsym}
\usepackage{epsf}
\usepackage{graphicx}
\usepackage{amsmath} 
\usepackage{amsfonts} 
\usepackage[latin1]{inputenc}
\usepackage[T1]{fontenc}

\newcommand{\etal}{et al.}

\newcommand{\Ki}{[\text{K}]_{\text{i}} }
\newcommand{\Ai}{[\text{A}]_{\text{i}} }
\newcommand{\Nai}{[\text{Na}]_{\text{i}}}
\newcommand{\Cai}{[\text{Ca}]_{\text{i}}}
\newcommand{\Caitot}{[\text{Ca}]_{\text{i}}^{\text{tot}}}
\newcommand{\SSspace}{\text{SS}}
\newcommand{\NSR}{\text{NSR}}
\newcommand{\SR}{\text{SR}}
\newcommand{\JSR}{\text{JSR}}
\newcommand{\myo}{\text{myo}}
\newcommand{\VSS}{v_{\SSspace}}
\newcommand{\Vmyo}{v_{\myo}}
\newcommand{\VJSR}{v_{\JSR}}
\newcommand{\VNSR}{v_{\NSR}}
\newcommand{\VSR}{v_{\SR}}

\newcommand{\CaNSR}{[\text{Ca}]_{\NSR}}
\newcommand{\CaJSRtot}{[\text{Ca}]_{\JSR}^{tot}}
\newcommand{\CaSStot}{[\text{Ca}]_{\SSspace}^{tot}}

\newcommand{\Ko}{[\text{K}]_{\text{o}}}
\newcommand{\Nao}{[\text{Na}]_{\text{o}}}

\newcommand{\dV}{\frac{dV}{dt}}
\newcommand{\iC}{\frac{1}{C_m}}
\newcommand{\Istim}{I_{\text{stim}}}
\newcommand{\Ic}{I_{\text{c}}}
\newcommand{\dK}{\frac{d\Ki}{dt}}
\newcommand{\Bi}{[\text{B}]_{\text{i}}}

\newcommand{\SI}{[\text{S}]_{\text{I}}}

\newcommand{\Sm}{[\text{S}]_{\text{m}}}
\newcommand{\Ei}{[\text{E}]_{\text{i}}}

\newcommand{\Cl}{\text{Cl}^-}
\newcommand{\Ca}{\text{Ca}^{2+}}
\newcommand{\K}{\text{K}^+}
\newcommand{\Na}{\text{Na}^+}
\newcommand{\dE}{\frac{d\Ei}{dt}}

\begin{document} 
 




\begin{center}
{\Large {\bf How to formulate membrane potential in a spatially homogeneous myocyte model?}}\\ 
\vspace{1cm} {\large \bf A.J.\ Tanskanen${}^{1,2}$\,},
{\large \bf E.I.\ Tanskanen${}^{3}$\,},
{\large \bf J.L.\ Greenstein\,${}^{1,2}$} and
{\large \bf R.L.\ Winslow\,${}^{1,2}$}\\
\vspace{0.3cm} {${}^1$}{\em The Center for Cardiovascular Bioinformatics and Modeling, %
The Johns Hopkins University School of Medicine and %
Whiting School of Engineering %
Baltimore MD 21218, USA}\\
\vspace{0.3cm} {${}^2$}{\em The Whitaker Biomedical Engineering Institute,
The Johns Hopkins University School of Medicine and %
Whiting School of Engineering, %
Baltimore, MD 21218, USA}\\
\vspace{0.3cm} {${}^3$}{\em Extraterrestrial Physics, NASA/Goddard Space Flight Center, 
Greenbelt, MD 20770, USA}\\
\end{center}

\begin{abstract}
\noindent Membrane potential in a mathematical model of a cardiac myocyte can be formulated in different ways.
Assuming a spatially homogeneous myocyte that is strictly charge-conservative and electroneutral as a whole, 
two methods will be compared:
(1) the differential formulation $dV/dt=-I/C_m$ of membrane potential used traditionally; and (2)
the capacitor formulation, where membrane potential is defined algebraically by the capacitor equation $V=Q/C_m$. 
We examine the relationship between the formulations, assumptions under which each formulation
is consistent, and show that the capacitor formulation provides a transparent, 
physically realistic formulation of membrane potential,
whereas use of the differential formulation may introduce unintended and undesirable behavior, 
such as monotonic drift of concentrations.
We prove that the drift of concentrations in the differential formulation arises as a compensation for
failure to assign all currents in concentrations.
As an example of these considerations, we present an electroneutral, explicitly  
charge-conservative formulation of Winslow~\etal~model (1999), and
extend it to describe membrane potentials between intracellular compartments.
\end{abstract}
 

\section{Introduction}
The single most important variable in an electrophysiological whole-cell model is membrane potential, 
defined as the potential difference across the cell membrane. 
It drives both gating of ion channels and fluxes of ions through the membrane.
The basis for electrophysiological single-cell modeling was first provided by 
the Hodgkin-Huxley model of the neuronal action potential (1952).
In their model, membrane potential was defined by an ordinary differential equation in which 
the time-derivative of membrane potential equals the sum of all ion currents through the membrane divided 
by membrane capacitance.  
Their model assumed constant ionic concentrations and dynamic concentrations 
were added in later models (DiFrancesco and Noble, 1985).
While tracking of concentrations made the models more realistic, doing so
introduced new problems, including drift of concentrations under repeated stimuli (see Fig. \ref{fig:figure1}), 
over-determined initial conditions and an infinite number of steady states
(Guan~\etal, 1997; Hund~\etal, 2001; Kneller~\etal, 2002). 

Long-term drift of concentrations is present in the DiFrancesco-Noble (Guan~\etal, 1997) and
Luo-Rudy models (Luo and Rudy, 1994).
Using numerical simulations, Hund et al. (2001) showed that the Luo-Rudy model has a steady state 
and no concentration drift, when it is assumed that the externally applied stimulus current is carried by potassium ($K^+$)
ions and that this ion flux contributes to the rate of change in intracellular $K^+$ concentration. 
Similarly, Kneller~\etal~(2002) demonstrated that this finding also holds true in their model.
However, neither the reason for nor the origin of ion concentration drift has been addressed quantitatively.
In this study, we explain the mechanism of concentration drift (Sec. \ref{sec:s31}) 
and examine the number of steady states present in the above models (Sec. \ref{sec:steady}).

We also investigate alternative formulations of membrane potential which preserve charge-conservation
and electroneutrality.
In particular, we compare what we refer to as the differential and capacitor formulations of membrane potential 
(Sec. \ref{sec:s2}), and consider issues affected by the formulation of membrane potential (Sec. \ref{sec:s3}).
Two examples (Sections \ref{sec:case1} and \ref{sec:case2}) show how 
membrane potential may be formulated rigorously in a computational model of the cardiac ventricular myocyte 
developed by Winslow et al. (1999; abbreviated WRJ). 
Our results show that the capacitor formulation provides a transparent, well-defined formulation of
membrane potential in a spatially homogeneous myocyte model.

\section{The myocyte as a capacitor}
\label{sec:s2}
\subsection{Formulation of membrane potential}
\label{sec:Vformulation}
Following Hodgkin and Huxley (1952), a spatially homogeneous single-cell model is described
as a parallel RC-circuit model.
Given an initial value, membrane potential $V$ is then defined by the differential equation 
\begin{equation}
\label{eq:HHV}
\frac{dV}{dt}=-\frac{1}{C_m}I,
\end{equation}
where $I$ is the sum of all outwardly directed membrane currents and $C_m$ is total membrane capacitance.
This formulation of membrane potential is referred to as the {\em differential formulation}. 

Membrane potential can be defined by assuming that a cell is a capacitor, 
consisting of an ideal conductor representing the interior of the cell, a dielectric representing 
the ability of the cell membrane to separate charge and a second ideal conductor representing the extracellular space
surrounding the cell. 
In this {\em capacitor formulation}, membrane potential is defined as the
ratio of charge $Q$ to total membrane capacitance $C_m$:
\begin{equation}
\label{eq:capacitor}
V=Q/C_m.
\end{equation}
Charge $Q$ includes contribution from all charged particles in a cell.
More specifically, membrane potential of a model containing a single intracellular compartment 
with $N$ ion species $\{S_k\}_{k=1}^N$ is given by
\begin{equation}
\label{eq:genV}
V=\frac{F\Vmyo}{C}\sum_{k=1}^{N}z_k[S_k]_i,
\end{equation}
where $z_k$ is the valence of species $S_k$,
$\Vmyo$ is volume of myoplasm and $F$ is Faraday's constant.

A biophysically detailed model often describes a myocyte using more than two compartments.
The capacitor formulation (\ref{eq:capacitor}) is extended to multiple compartments
by describing the cell as a network of capacitors (Fig. \ref{fig:intracellular}A), 
where each dielectric corresponds to an interface between two compartments. 
The interior of a compartment is considered to be an ideal conductor insulated from other compartments by a membrane. 
The membrane potentials can be expressed as functions of charges of compartments
and of capacitances of interfaces.
In particular, a compartment that is completely enclosed within a larger compartment
influences the membrane potential of the surrounding compartment only via charge, 
geometry is irrelevant (Griffiths, 1989).
For example, Figures \ref{fig:intracellular}A and \ref{fig:intracellular}B show graphical 
and capacitor representations of a three-compartment cell model. 
Membrane potentials are given by $V_1=(Q_{\text{myo}}+Q_{\text{mito}})/C_1$, 
and $V_2=Q_{\text{mito}}/C_2$,
each depending only on net charge of the enclosed volume.

\subsection{Electroneutrality and homogeneous concentrations}
\label{sec:electroneutrality}
Membrane potential is generated by a small number of ions, the bulk ionic concentration is
electroneutral (Hille, 2001).
We require that (1) a model is electroneutral as a whole, that is, the net charge is zero;
and that (2) the concentrations are spatially homogeneous (i.e. well-mixed) in each compartment.
Since such requirements are not necessarily satisfied in the presence of charged particles, 
we will show this to be the case for the model presented here.

In an ideal conductor, induced charges balance electric field 
in such a way that potential is constant inside the conductor. 
Hence the interior of each compartment is exactly electroneutral in the capacitor approximation
and all net charge is located on the compartment boundary.
This is consistent with narrow physiological range of membrane potentials, which also requires
that sum $\sum z_k[S_k]_i$ must be nearly zero. 

Both anions and cations must be present in each compartment, since
the absence of anions (cations) would imply that all cations (anions) are located on the membrane.
The concentration of ions that comprise the induced charge is extremely low compared to bulk concentrations and
can typically be neglected.
For example, in the WRJ model, range $[-100,50]$ mV limits induced charge to correspond to 
a monovalent ion concentration in range $[-0.5,0.25]$ $\mu\text{mol}/\text{L}$, which is negligible compared to
most intracellular ion concentrations.
Electroneutrality of the interior allows the assumption that ion concentrations are homogeneous in each compartment. 

In a typical experimental setup, a myocyte is embedded in an electroneutral solution of
essentially infinite volume relative to the cell volume.
While this extracellular space (ES) is treated like any other compartment in a cell model, 
the assumption of infinite volume (implied by constant extracellular concentrations)
brings certain extra complications that will be addressed in the following.
If the myoplasm has charge $Q$, charge of the ES is $-Q$. 
However, if the charge density of ES, $F\sum_{k=1}^{N}z_k[S_k]_e$ (notation as in (\ref{eq:genV})),
is non-zero, then the ES has infinite charge which implies infinite membrane potential. 
Hence, the ES must have {\em zero} charge density, but typically {\em non-zero} charge.
The paradox arises from infinite volume: if the charge of
the ES is finite, the average charge density is zero because
no finite flow of ions can change the concentrations within an infinite volume eventhough charge is altered. 
As with all other compartments, all net charge is located on the boundary of the ES, 
and the remainder of the compartment is exactly electroneutral. 
In conclusion, the capacitor approximation of the cell is consistent 
with the requirement of electroneutrality and the assumption that ions
are homogeneously distributed in a compartment.

\subsection{Connection between the formulations}
\label{sec:s21}
Next, we will study the exact connection between the two formulations of membrane potential.
In the differential formulation, membrane potential is defined as an independent variable and
does not depend on ion concentrations, 
whereas in the capacitor formulation, membrane potential is a function of concentrations.
Hence, in the differential formulation initial conditions are over-determined,
since initial conditions are assigned independently for interdependent variables.
This issue can be resolved by introducing implicitly-defined ion concentrations 
in the differential formulation, as will be shown in the following.

For example, assume a one intracellular compartment model with $M$ concentrations $\{\Sm\}_{m=1}^M$ 
and $N$ currents $\{I_k\}_{k=1}^N$, is defined in the differential formulation by
\begin{equation}
\label{eq:no5}
\begin{aligned}[2]
&\dV=-\sum_{k=1}^NI_k/C_m,\\
&\frac{d}{dt}\Sm=-\sum_{l=1}^{N_m}I_{a_{m,l}}/(z_mFv),
\end{aligned}
\end{equation}
where $v$ is the volume of the compartment, $a_{m,l}, l=1,...,N_m,$ indexes the $N_m$
currents assigned to ion species $\text{S}_m$ of valence $z_m$.

Assume that not all currents carry ion species modeled by any of the intracellular concentrations $\Sm$.
Such current are assigned to an implicitly-defined monovalent concentration $\SI$, 
\begin{equation}
\label{eq:no6}
\frac{d}{dt}\SI=-\sum_{l=1}^{N_I}I_{a_{I,l}}/Fv,
\end{equation}
where $N_I=N-\sum_mN_m$. Combining equations (\ref{eq:no5}) and
(\ref{eq:no6}) yields
\begin{equation}
\label{eq:7}
\frac{d}{dt}V=Fv\bigl(\sum_{m}\frac{d}{dt}\bigl(z_m\Sm\bigr)+\frac{d}{dt}\SI\bigr)/C_m.
\end{equation}
Integrating equation (\ref{eq:7}) gives membrane potential
\begin{equation}
\label{eq:SiV}
V=Fv\bigl(\sum_{m}z_m\Sm+\SI\bigr)/C_m.
\end{equation}
The initial value of $\SI$ is determined by the assumption regarding initial charge density.
Equation (\ref{eq:SiV}) shows the exact connection of 
the differential formulation (\ref{eq:no5}) to the capacitor formulation (\ref{eq:capacitor}) 
(in the two-compartment case, but the derivation can be generalized to any number of compartments).
The major difference between the formulations is $\SI$, which accounts for
both the currents missing from equation (\ref{eq:no5}) and the charge missing from
initial conditions in the differential formulation. 
The formulations are equivalent if $\SI$ is constant and all movement of charge is captured by the currents.
Concentration $\SI$ and its time evolution are the source
of a variety of problems, as will be shown in the following.

\section{Consequences of membrane potential formulation}
\label{sec:s3}
In this section, we will show how the physical constraints presented above
influence the formulation of a cell model.

\subsection{Concentrations and drift}
\label{sec:s31}
Here we examine how and under what conditions 
spurious concentration drift arises in the differential formulation.
Assume a model with $\K$ concentration $\Ki$ and a constant anion concentration $\Bi$, in which
membrane potential is given by 
\begin{equation}
F\Vmyo(\Ki-\Bi)/C_m.
\end{equation}
It is clear that stimulus current $\Istim$ must influence $\Ki$ if it is to modify membrane potential.
Charge-conservation and electroneutrality require that 
the stimulus current originates from the ES.

In the differential formulation, stimulus current has not historically been
assigned to a particular ion concetration. 
Formulating the above model in this manner yields
a model comparable to the one defined above defined by
\begin{equation}
\label{eq:Vforsimple_diff}
\begin{aligned}[2]
\dV&=-\iC(I_K+\Istim),\\
\dK&=-\frac{1}{Fv_{myo}}I_K,
\end{aligned}
\end{equation}
where $I_K$ is $\K$ current through the cell membrane. 
Note that concentration $\Bi$ does not appear in (\ref{eq:Vforsimple_diff}).
If we assign the stimulus current to implicitly-defined anion concentration $\Ei$ 
(no information on valence of species $E$ is contained in equation (\ref{eq:Vforsimple_diff})), 
\begin{equation}
\label{eq:defE}
\dE=\frac{1}{Fv_{myo}}\Istim,
\end{equation}
and equation (\ref{eq:SiV}) implies that
\begin{equation}
\label{eq:VforKE}
V=F\Vmyo(\Ki-\Ei)/C_m.
\end{equation}
Since all currents are explicitly accounted for in the concentrations, equation (\ref{eq:VforKE})
shows that the model defined by equation (\ref{eq:Vforsimple_diff}) is charge-conservative, 
when concentration $\Ei$ is included in the formulation.
A positive stimulus current decreases $\Ei$ monotonically and eventually makes it negative, 
that is, $\Ei$ has spurious drift.
To keep membrane potential in the physiological range, the decrease of $\Ei$ must be compensated
for by a decrease of $\Ki$. 
Drift occurs in both the capacitor (equation \ref{eq:VforKE}) and differential formulations (equation \ref{eq:Vforsimple_diff}) 
of the model, however, the cause of drift is transparent only in the capacitor formulation.

Explanation of the drift presented above may be generalized to any number of concentrations. 
In a model with unassigned current $\Ic$, membrane potential is given by equation (\ref{eq:SiV}).
The current $\Ic$ either increases or decreases $\SI$ monotonically. 
To keep membrane potential in the physiological range, charge density $\sum_{m\neq I}z_m\Sm$, must compensate
for the change in $\SI$,
which is manifested as a monotonic drift of at least one of the concentrations $\Sm$.
Even a minor drift in a single ion concentration will influence membrane potential if it goes uncompensated.
Hence drift is inevitable if there exists a current that transports charge and hence affects membrane 
potential, but does not contribute to any ion concentration. This is consistent with previous numerical
studies (Hund et al., 2001; Kneller et al., 2002).
Previously presented descriptions of the drift (e.g., Hund~\etal~(2001)) have stated that lack of charge conservation
is at the root of the problem but did not demonstrate how drift arises. 
As shown above, concentration drift is a compensation for
the change in the implicitly-defined concentration $\SI$.

How does the drift depend on pacing rate?
Since drift is relative to the charge transported by the stimulus current, 
higher pacing rate leads to more rapid decrease of $\Ei$, and consequently to faster drift in $\Ki$, 
consistent with numerical simulations of Hund~\etal~(2001).

\subsection{Steady state and time scales}
\label{sec:steady}
An infinite number of steady states were observed in simulations of Guan et al. (1997), 
when initial concentrations were varied. 
If voltage is kept constant but concentrations are changed in the differential formulation,
concentration $\SI$ (equation (\ref{eq:SiV})) is changed (Sec. \ref{sec:s31}). 
If $\SI$ remains constant during a simulation, different values of $\SI$ result in different steady states. 
Hence, an infinite possible number of initial values of $\SI$ yield an infinite number of steady states.
On the other hand, $\SI$ is not constant if any of the currents implicitly assigned to $\SI$ is non-zero, 
in which case the model may not have a steady state. In the capacitor formulation,
$\SI$ is explicitly included in the initial conditions, which resolves this issue.

The time scale of exponential relaxation to steady state can be described 
by time constant $\tau$, measured by fitting $A+Be^{-t/\tau}$
to either the diastolic voltage, $\Na$ or $\K$ concentrations (Fig. 3).
The time constant is mainly determined by the balance of $\Na$ and $\K$ concentrations.
This time constant divides the model time scales into two regimes: slow ($t>\tau$) and fast ($t<\tau$). 
The fast times scale is on the order of a few seconds or less, whereas the slow time scale is on the order
of tens or hundreds of seconds.
In a study of fast time scale processes, $\Na$ and $\K$ concentrations can be clamped. 
However, if concentrations are clamped for a time period comparable
to $\tau$, incorrect behavior will occur since implicit concentrations will change significantly during the simulation. 
Analogously, in voltage clamp a control current holds the membrane potential at the desired level. 
If this current is not assigned to any concentration, it is accounted for in
implicit concentration and the behavior of the system will be incorrect
in protocols with duration comparable to the time constant.

\subsection{Rapid equilibrium approximation}
\label{sec:rea}
The rapid equilibrium approximation (REA) often provides a powerful method to simplify a cell model
(see, e.g., Hinch et al. (2004)). 
However, the REA requires instantaneous movement of charge in the model.
Due to currents associated with instantaneous movement of charge, the differential and capacitor formulations 
may yield different results, as will be shown in the following.

Assume a model consists of a small-volume intracellular compartment and 
an infinite-volume extracellular compartment which are connected by 
a single ion channel that is described as a two-state Markov model. 
The intracellular compartment has anion concentration $\Ai$ equal to 
extracellular $\K$ concentration $\Ko$. 
Also assume that the intracellular $\K$ concentration $\Ki$ is an instant function of the state of the channel 
(essentially a REA): if the channel is closed $\Ki=0$, otherwise $\Ki=\Ko$. 
Membrane potential $V=F(\Ki-\Ko)/C_m$ depends on state of the ion channel.
Assuming that $\alpha$ and $\beta$ are opening and closing rates, respectively, of the
ion channel, the probability $P_C$ that the ion channel is closed evolves according to 
equation $dP_C/dt=\beta-(\alpha+\beta)P_C$.

Assume a large ensemble of $N$ intracellular compartments 
described above to be located within a single cell.  
According to capacitor formulation, membrane potential of the cell is then $V(t)=-NFP_C(t)\Ko/C_m$.
Since each individual channel allows non-zero flux only at the time of channel closing and opening,
the total current seems to mostly be zero. However, the ensemble current through the cell membrane 
due to REA is $I=F(\beta-(\alpha+\beta)P_C)\Ko N$.
This current is 
missed in the naive application of the differential formulation, which thus yield incorrect membrane potential.

In a recent reformulation of the WRJ model by Greenstein et al. (2005, submitted to Biophys. J.),
the REA was employed.
In this case, current due to the REA is not easily solved 
from the equations, and use of the differential formulation introduces systematic,
albeit small, error in calculation of the membrane potential (Fig. \ref{fig:figure1}D-E).
The small magnitude of this error is due to the small ratio of the diadic volume to the total cell volume.
When a similar approximation is applied to a larger compartment, 
the difference can be significant.

\section{Case study 1: the WRJ model}
\label{sec:case1}
The WRJ model describes intracellular $\Ca$ dynamics of the cardiac ventricular myocyte.
The model consists of four intracellular compartments: myoplasm, network sarcoplasmic reticulum (NSR), 
junctional sarcoplasmic reticulum (JSR) and the diadic space (SS). The myocyte is
embedded in the extracellular space containing constant ion concentrations.
The WRJ model has concentration drift and consequently does not have a steady state; it 
does not exhibit physiologically realistic dependence of APD on pacing rate, nor is it electroneutral. 
As an application of the considerations in the previous two sections, 
we reformulate the WRJ model and address these issues in the following.

To address the above mentioned issues, we modified the model of (Mazhari~\etal, 2002) which was based on the WRJ model
in the following manner:
stimulus current $I_{\text{stim}}$ is assigned to intracellular $\K$ concentration $\Ki$; 
conductances of background calcium current $I_{\text{Ca,b}}$, 
background $\Na$ current $I_{\text{Na,b}}$ and $\Na$-$\K$ pump $I_{\text{NaK}}$ 
were adjusted to balance concentrations in the long term; and 
$\K$, $\Na$ and stationary anion concentrations were added to compartments if not already present. 
We reformulated membrane potential using the capacitor formulation, in which membrane
potential is given by 
\begin{equation}
\label{eq:V_WRJ}
V=[Q_i+Q_{\text{NSR}}+Q_{\text{JSR}}+Q_{\text{SS}}]/C_m,
\end{equation}
where charges $Q_k$ are defined through
\begin{equation}
\begin{split}
&Q_i=F\Vmyo(\Ki+\Nai+2\Caitot-[S]_i),\\
&Q_{\text{NSR}}=F\VNSR([\text{K}]_{\NSR}+[\text{Na}]_{\NSR}+2\CaNSR-[\text{S}]_{\NSR}),\\
&Q_{\text{JSR}}=F\VJSR([\text{K}]_{\JSR}+[\text{Na}]_{\JSR}+2\CaJSRtot-[\text{S}]_{\JSR}),\\
&Q_{\text{SS}}=F\VSS([\text{K}]_{\SSspace}+[\text{Na}]_{\SSspace}+2\CaSStot-[\text{S}]_{\SSspace}),
\end{split}
\end{equation}
where $v_k$ is volume of compartment $k$, 
concentration $[S]_k$ is static anion concentration in compartment $k$.
The index {\em tot} refers to the sum of buffered and free $Ca^{2+}$.

\begin{figure}
\includegraphics[width=11cm]{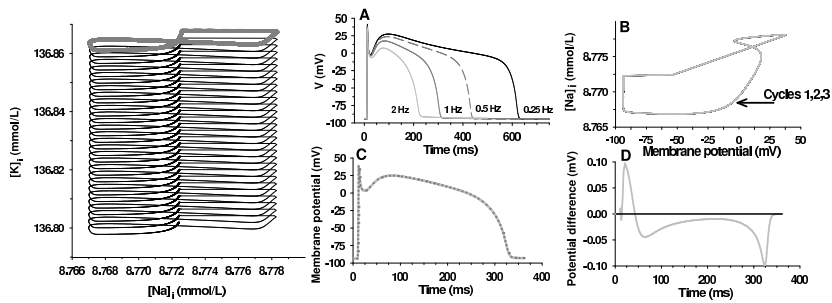}
\caption{(A) When stimulus current is assigned to $\K$ concentration in Case Study 1 model,
the model has a limit cycle ("steady state"; grey cycle) in $(\Ki,\Nai)$ phase space.
When the stimulus current is not assigned to any concentration, the model exhibits 
monotonic drifts away (black solid line) from the steady state;
(B) Steady state action potentials at 0.25-2 Hz pacing rates in the Case Study 1 model. 
Membrane potential (ordinate; mV) is plotted against time (abscissa; ms);
(C) Three simulations started from three different initial conditions (each with zero net charge)
all approach the same steady state limit cycle in $(\Nai,V)$ phase space in the Case study 1 model. Sodium
concentration (ordinate; mmol/L) is plotted against membrane potential (abscissa; mV);
(D) Action potentials simulated using the differential (solid grey line) and capacitor formulations
(dashed dark grey line) and (E) the potential difference between the formulations in 
Greenstein et al. (2005, submitted to Biophys. J.) model.}
\label{fig:figure1}
\end{figure}

\begin{figure}
\includegraphics[width=8cm]{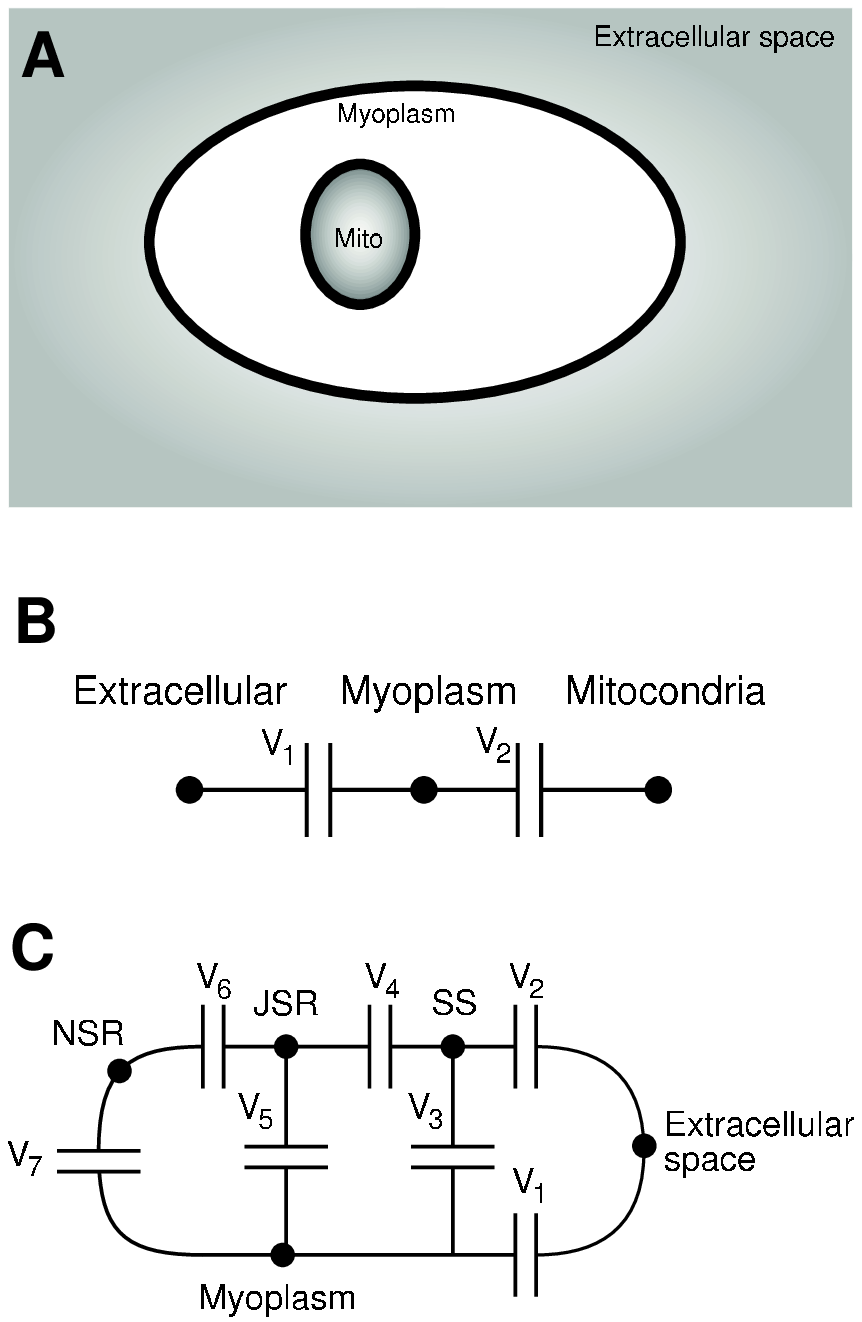}
\caption{({\em A}) Spatial representation and ({\em B}) capacitor
 representation of a myocyte model with three compartments: extracellular space, 
 myoplasm and mitocondria;
(C) Capacitor representation of five-compartment single-cell myocyte model, consisting of
myoplasm, subspace (SS), junctional sarcoplasmic reticulum (JSR), network sarcoplasmic reticulum (NSR) 
compartments and an extracellular space. Membrane potentials between the compartments 
are denoted by $V_k$.}
\label{fig:intracellular}
\end{figure}

In a ventricular myocyte, action potential duration (APD) depends on pacing rate. 
Figure \ref{fig:figure1}A shows steady state action potentials 
at pacing rates 0.25-2 Hz, with higher pacing rates producing shorter action potentials.
The pacing rate dependence of APD arises mainly as a result of long-term changes in $\Nai$ and $\Ki$. 
Figure \ref{fig:figure1}B shows simulations started 
(1) with intracellular concentrations set to extracellular concentrations;
(2) from a minimally perturbed steady state (static charges unchanged);
and (3) with steady state $\Nai$ and $\Ki$ replaced by $\Nai+1$ and $\Ki-1$.
In each case, limit cycles in $(V,\Nai)$ phase space are identical within numerical accuracy,
showing that the steady state is unique, given parameters including pacing rate. 

\begin{figure}
\includegraphics[width=8cm]{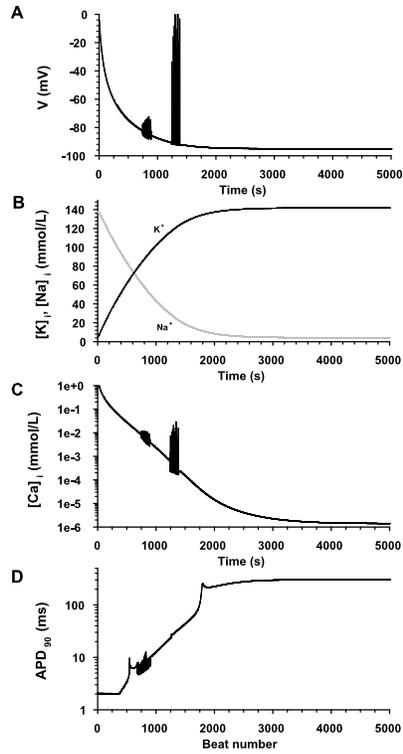}
\caption{
A Case study 1 simulation run demonstrating how a myocyte makes its way from a state with no concentration
gradients to physiological steady state:
(A) Resting membrane potential (ordinate; mV) sampled once a second,
plotted against time (abscissa; s); (B) Sodium (gray) and Potassium (black) concentrations
(ordinate; mmol/L) against time (abscissa; ms);
(C) Diastolic calcium concentration (ordinate; mmol/L; logarithmic scale) plotted against time (abscissa; ms); 
(D) Action potential duration (ordinate; ms; logarithmic scale) plotted against time (abscissa; ms).}
\label{fig:start_zero}
\end{figure}

Figure \ref{fig:start_zero} illustrates long-term changes in the model, sampled at 1 Hz. 
The simulation is started from a state with no concentration gradients,
Figure \ref{fig:start_zero}A shows diastolic membrane potential.
Figure \ref{fig:start_zero}B exhibits the increase of $\Nai$ (initially set to $\Nao$) 
and the decrease of $\Ki$ (initially set to $\Ko$) to their steady state values.
Homeostasis is approached approximately exponentially with a time constant of 90 seconds.
Figure \ref{fig:start_zero}C shows the decrease of diastolic $\Cai$ towards its steady state.
Two oscillatory regimes of $\Cai$ emerge:
first at roughly 800 seconds with small, subthreshold oscillations;
second with large amplitude oscillations at roughly 1,400 seconds.
Figure \ref{fig:start_zero}D shows the non-trivial time evolution of APD during the simulation. 

\section{Case study 2: Compartmental membrane potentials}
\label{sec:case2}
Intracellular compartments are important for proper myocyte function.
In particular, the sarcoplasmic reticulum (SR) stores $\Ca$ ions.
While the process of uploading of $\Ca$ into the SR is electrogenic, it does not seem to 
influence SR membrane potential that is observed to be small in amplitude ($\VSR$; Bers, 2001).
Pure diffusion of ions through SR membrane cannot explain $\VSR$, 
since it would balance ion species separately without consideration to $\VSR$. 
A small $\VSR$ requires that movement of counter-ions, likely $\K$ and $\Cl$, balances the potential difference
generated by $\Ca$ movement (Pollock~\etal, 1998; Kargacin~\etal, 2001). 
Indeed, SR membrane is known to have $\Cl$ channels gating according to $\VSR$
(Townsend and Rosenberg, 1995), suggesting that $\VSR$ does affect
$\Ca$ handling unless kept at almost zero voltage.
Somewhat counter-intuitively, zero $\VSR$ requires currents driven by $\VSR$.

\subsection{Formulation}
To better understand the basis for $\VSR$ and the concentrations of ions in SR, 
we extend the model of Case study 1 to describe intracellular membrane potentials. 
In this case, the cell is described as a network of capacitors shown in Fig. \ref{fig:intracellular}B.
Membrane potential $V_k$ of interface $k$ (Fig. \ref{fig:intracellular}B) can be expressed
as a function of capacitance $C_k$ of and charge $q_k$ bound to interface $k$ by
\begin{equation}
V_k=q_k/C_k.
\end{equation}
Given net charges $Q_m=Fv_m\sum_lz_l[S_l]_m$, the charges $q_k$ are 
\begin{equation}
\begin{split}
q_1&=\frac{-\beta (b+C_5s+Q_{\JSR})-Q_{\NSR}/C_7-s}{r+\beta(1+C_3\gamma+C_5r)},\\
q_2&=-q_1-Q_e,\\
q_3&=q_1C_3/C_1-q_2C_3/C_2,\\
q_4&=q_2-q_3-Q_{\SSspace},\\
q_5&=q_3C_5/C_3-q_4C_5/C_4,\\
q_6&=Q_{\JSR}-q_4+q_5,\\
q_7&=-q_6-Q_{\NSR},\\
\end{split}
\end{equation}
where $\beta=\frac{1}{C_6}+\frac{1}{C_7}, 
\gamma=\frac{1}{C_1}+\frac{1}{C_2},
b=Q_{\SSspace}+Q_{e}(1+\frac{C_3}{C_2}),
r=[1+\gamma(C_3+C_4)]/C_4,
s=(Q_e(C_4C_2+C_2C_4+C_3)+C_2Q_{\SSspace})/(C_4C_2).$
Charge $Q_e$ is given by electroneutrality, $Q_e=-(Q_m+Q_{\JSR}+Q_{\NSR}+Q_{\SSspace})$.

An ion channel senses voltage $V$ across the membrane in which the channel is located. 
Then the flux $J_G$ of ions of species $S$ between compartments $a$ and $b$ is
given by the Goldman equation (Hille, 2001)
\begin{equation}
\label{eq:GHK}
  J_{G}=D\frac{zVF}{RT}\frac{[S]_{a}e^{zVF/RT}-[S]_{b}}{e^{zVF/RT}-1}.
\end{equation}
where $D$ is the diffusion coefficient, $z$ valence, 
$T$ temperature and $R$ universal gas constant.
Ion channels (other than $\Ca$ channels) between intracellular compartments are assumed to
always be open and to be selective to a single ion species.
In particular, myoplasm-SS and JSR-NSR interfaces have large-conductance, permanently-open pores. 
Each compartment contains $\Cl$ and stationary anion concentrations. 

To include the effect of voltage $V_7$ modulating $\Ca$ transport to NSR, 
we derived a model for the SERCA2a pump assuming 
Michaelis-Menten kinetics, that SERCA2a achieves equilibrium with $\Ca$ instantaneously,
and that rates $\alpha$ and $\beta$ out of the intermediate Michaelis-Menten state
are related by $e^{-\eta\Delta G/kT}=\alpha/\beta$, where $\eta\Delta G$ is free energy of ATP breakdown.
The flux through SERCA2a is 
\begin{equation}
\label{eq:SERCA}
J_{\text{SERCA}}=V_{max}\frac{\Cai^2/K_i^2-\CaNSR^2/K_{NSR}^2}{1+\Cai^2/K_i^2+\CaNSR^2/K_{NSR}^2},
\end{equation}
where $K_{NSR}=K_ie^{-(\eta\Delta G_{ATP}-4e_0V_7)/kT}$, in which $\eta=0.74$,
$e_0$ is elementary charge, $K_i=300$ nmol/L.

\subsection{Results}
\begin{figure}
\includegraphics[width=6cm]{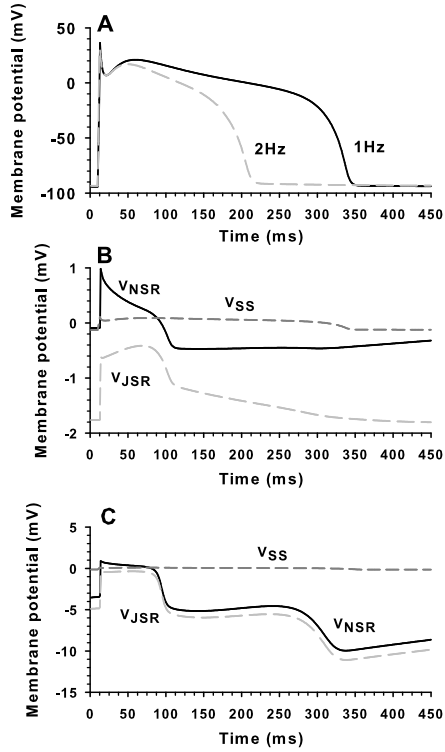}
\caption{Results from the Case study 2 model: (A) Shape of baseline action potential, paced at 1 Hz (solid black) 
and at 2 Hz (dashed grey). Membrane potential (ordinate; mV) is 
plotted against time (abscissa; ms);
(B) Baseline membrane potentials (ordinate; mV) on myoplasm-JSR, myoplasm-NSR 
and myoplasm-SS interfaces plotted against time (abscissa; ms).
(C) Increased membrane potentials (ordinate; mV) on myoplasm-JSR, myoplasm-NSR 
and myoplasm-SS interfaces plotted against time (abscissa; ms) in a simulation with 
reduced diffusion between intracellular compartments.}
\label{fig:case2_V}
\end{figure}

The model has a realistic dependence of APD on pacing rate, as evidenced by Figure \ref{fig:case2_V}A.
Intracellular membrane potentials are consistently small but non-zero (Fig. \ref{fig:case2_V}B).
Membrane potentials $V_3$ and $V_7$ between myoplasm and SS and NSR are essentially zero.
Membrane potentials $V_4, V_5$ and $V_6$ between JSR and other compartments are non-zero, 
roughly 1 mV, as a result of $\Ca$ buffering by Calsequestrin in JSR, 
the absence of which would reduce these membrane potentials to essentially zero.
External membrane potential of SS ($V_2$) tracks
closely external membrane potential of myoplasm, $V_1$. 

When diffusion rates between myoplasm and SR are reduced,
intracellular membrane potentials are increased (Fig. \ref{fig:case2_V}C). 
The maximal NSR membrane potential that SERCA can overcome by employing energy of ATP hydrolysis (56 kJ/mol; Bers, 2001)
is 237 mV. 
Without the movement of counter-ions, $7.5\ \mu$mol/L would be the maximal increase of $\CaNSR$
over $\Cai$, while the measured concentration difference is roughly 0.7 mmol/L (Bers, 2001).
Since cells are sensitive to any disruption in $\Ca$ handling, even a minor SR membrane potential has functional consequences.

Case study 2 demonstrates that (1) Intracellular membrane potentials can be incorporated in a cell model,
and that (2) they give physical constraints on intracellular concentrations and 
require delicate balance of charges;
(3) the magnitude of SR membrane potential $\VSR$ can be explained by movement of counter-ions;
and (4) buffering of $\Ca$ affects JSR membrane potential.

\section{Discussion}
\label{sec:discussion}
\subsection{Which formulation is appropriate?}
The capacitor formulation expresses membrane potential as a function of charge and capacitance,
whereas the differential formulation uncouples membrane potential from concentrations.
The main differences between the two formulations are
an integration constant (the differential formulation is time derivative of the capacitor equation),
and "independence" of membrane potential from concentrations in the differential formulation.
These two issues imply the presence of a dynamic, implicitly-defined ion concentration(s)
in the differential formulation (Sec. \ref{sec:s31}), which makes interpreting 
simulation results difficult and prone to errors, in addition to the presence of
spurious drift of concentrations and issues with steady state. 
The differential formulation is equivalent to a particular capacitor formulation, 
which, however, may not be the one intended due to the presence of implicit concentrations.

The capacitor formulation requires "almost" electroneutrality and carefully chosen initial conditions
due to sensitivity of membrane potential to net charge. However, these are physical constraints, 
since even a small additional charge can have a drastic effect on membrane potential. 
The capacitor formulation provides a transparent formulation of membrane potential, 
and requires no implicitly-defined concentrations.
The differential formulation is best viewed as a shorthand notation for the more complete
and better-defined capacitor formulation.

\subsection{Comparison with previous studies}
Guan et al. (1997) show that the DiFrancesco-Nobel model has infinite number of steady states, when initial concentrations
are varied. To address this issue, they suggest that concentrations should be treated as parameters. 
However, they neglected the implied concentration in their study,
inclusion of which resolves the issue.
Consistent with our explanation (Sec. \ref{sec:s31}), Kneller et al. (2002) observed 
that if the sum of concentration changes in initial conditions is zero, the steady state stays the same. 

The sinoatrial node model by Endresen~\etal~(2000) uses the capacitor equation for membrane potential. 
However, the reasoning behind their definition is different from ours. 
The membrane potential of a simplified model described in Sec. \ref{sec:s31} is given by equation $F\Vmyo(\Ki-\Bi)/C_m$.
Our interpretation of this equation is that concentration $\Bi$ 
represents intracellular concentration of an anion concentration required for electroneutrality.
In Endresen~\etal~(2000), $F\Vmyo\Bi/C_m$ represents charge of the extracellular space, which
however cannot be computed directly from the concentrations in the extracellular space due 
to its infinite volume.

Varghese and Sell (1997) derived a "latent conservation law" showing that membrane potential can be expressed as
a function of concentrations. They showed that when each current is
assigned to contribute to some concentration, the implicit concentration is constant.
As they point out, the reason for emergence of the conservation law is best understood 
from the capacitor equation (\ref{eq:capacitor}). 
However, they did not consider time-dependent changes in the implicit concentration, study of which
is required to understand model phenomena such as concentration drift.

Using numerical simulations Hund~\etal~(2001) showed that assignment of stimulus current to $\Ki$ 
is sufficient to remove concentration drift in the Luo-Rudy model (Luo and Rudy, 1994). 
They interpret the drift as a consequence of "non-charge-conservative" formulation of the model. 
However, we proved that a "non-charge-conservative" formulation
is actually charge-conservative (Sec. \ref{sec:s21}), when the implicit concentrations is taken into account,
even in the presence of spurious concentration drift (Sec. \ref{sec:s31}). 
In particular, this means that the equations defining the differential formulation require that the system
is always charge-conservative.
On the other hand, a model can be "non-charge-conservation", 
which, however, does not necessarily indicate the presence of
concentration drift.
Hence, explanation of concentration drift requires study of charge densities implied by, but not explicitly
included in, the differential formulation (Sec. \ref{sec:s31}). Hund~\etal~(2001) further state
that the differential and capacitor formulations are always equivalent. 
However, that is true only if the currents present in the model capture all movement of charge, 
which is not necessarily the case in, e.g., rapid equilibrium approximation (Sec. \ref{sec:rea}).

In this article, 
we showed that the capacitor formulation provides a physically consistent, well-defined formulation
of membrane potential, and it avoids the problems all too often found in the differential formulation.
In conclusion, we see little reason to use the differential formulation as the 
definition of membrane potential in a spatially homogeneous myocyte model. 

\section*{Acknowledgements}
The cell models are available at website of the Center for Cardiovascular Bioinformatics and Modeling
(http://www.ccbm.jhu.edu/).
AT wishes to thank Dr. Reza Mazhari, Dr. Sonia Cortassa and Tabish Almas for helpful discussions.
This study was supported by NIH (RO1 HL60133, RO1 HL61711, P50 HL52307), 
the Falk Medical Trust, the Whitaker Foundation and IBM Corporation.
The work of ET was funded by the National Research Council and Academy of Finland.

\section*{References}

\begin{description}
\item[] Bers, D.M.,\ 2001. Excitation-Contraction Coupling and Cardiac Contractile Force, Second edition.
Kluwer Academic Publishers. Dordrecht.
\item[] DiFrancesco, D., Noble, D., 1985. A model of cardiac electrical activity incorporating ionic pumps and concentration changes. 
Phil. Trans. R. Soc. Lond. B. Biol. Sci. 307:353-398.
\item[] Endresen, L.P., Hall, K., H\o ye, J.S., Myrheim, J., 2000.
A theory for the membrane potential of living cells. 
Eur. Biophys. J. 29:90--103.
\item[] Griffiths, DJ., 1989. Introduction to electrodynamics, second edition. Prentice-Hall, Englewood Cliffs NJ.
\item[] Guan, S., Lu, Q., Huang, K. 1997. A discussion about the DiFrancesco-Noble model.
J. Theor. Biol. 189(1):27-32. 
\item[] Hille, B., 2001. Ion channels of excitable membranes, third edition.
Sinauer, Sunderland MA.
\item[] Hinch, R., Greenstein, J.L., Tanskanen, A.J., Xu, L., Winslow, R.L., 2004. 
A simplified local control model of calcium induced calcium release in cardiac ventricular myocytes.
Biophys. J. 87(6):3723-36.
\item[] Hodgkin, A.L., Huxley, A.F., 1952. A quantitative description of membrane current and its application to conduction and excitation in nerve. 
J. Physiol. (Lond.) 117, 500--544.
\item[] Hund, T.J., Kucera, J.P., Otani, N.F., Rudy, Y., 2001. Ionic charge 
concervation and long-term steady state in Luo-Rudy dynamic cell model. 
Biophys. J. 81, 3324--3331.
\item[] Kargacin, G.J., Ali, Z., Zhang, S.J., Pollock, N.S., Kargacin, M.E., 2001. Iodide and bromide inhibit $\Ca$ uptake by cardiac sarcoplasmic reticulum. 
Am. J. Physiol. Heart Circ. Physiol. 280(4):H1624-34
\item[] Kneller, J., Ramirez, R.J., Chartier, D., Courtemanche, M., Nattel, S., 2002. 
Time-dependent transients in an ionically based mathematical model of the canine atrial action potential.
Am. J. Physiol. Heart Circ. Physiol. 282: H1437--H1451.
\item[] Luo, C.H., Rudy, Y., 1994. A dynamic model of the cardiac ventricular action potential. I. Simulations of ionic currents and concentration changes.
Circ. Res. 74(6):1071--96.
\item[] Mazhari, R., Greenstein, J.L., Winslow, R.L., Marban, E., Nuss, H.B., 2001. 
Molecular interactions between two long-QT syndrome gene products, HERG and KCNE2, rationalized by in vitro and in silico analysis. 
Circ. Res. 89(1):33--38.
\item[] Pollock, N.S., Kargacin, M.E., Kargacin, G.J., 1998. 
Chloride channel blockers inhibit $\Ca$ uptake by smooth muscle sarcoplasmic reticulum. 
Biophys. J. 75:1759--1766.
\item[] Townsend, C., Rosenberg, R.L., 1995. Characterization of a chloride channel reconstituted from cardiac sarcoplasmic reticulum. 
J. Membr. Biol. 147(2):121-36.
\item[] Varghese, A., Sell, AR., 1997. A conservation principle and its effect on the formulation of Na-Ca exchanger current in the cardiac cells. 
J. Theor. Biol. 189, 33--40.
\item[] Winslow, R.L., Rice, J., Jafri, S., Marban, E., O'Rourke B., 1999. 
Mechanisms of altered excitation-contraction coupling in canine tachycardia-induced heart failure, II: model studies.
Circ. Res. 84(5):571-86. 
\end{description}

\newpage

\newpage

\newpage

\newpage

\end{document}